\documentclass[prb,aps,twocolumn,preprintnumbers,superscriptaddress]{revtex4-1}
\usepackage{latexsym,epsfig,amssymb,amsfonts,amsmath,graphicx,color,bm,times,hyperref,amstext,epstopdf,braket, float}
\usepackage{blindtext}
\usepackage{amssymb}
\usepackage{soul}
\usepackage{wasysym}
\providecommand{\qu}[1]{\textcolor{black}{#1}}
\bibpunct{[}{]}{,}{n}{}{}

\begin{document}

\title{Controlling valley splitting and Polarization of dark- and bi-excitons in monolayer WS$_2$ by a tilted magnetic field}

\author{Fanyao Qu}
\email{fanyao@unb.br}
\affiliation{Instituto de F\'{\i}sica, Universidade de Bras\'{\i}lia, Bras\'{\i}lia-DF 70919-970, Brazil}
\author{Helena Bragan\c{c}a}
\affiliation{Departamento de F\'isica, Universidade Federal de S\~{a}o Carlos, Brazil}
\affiliation{Instituto de F\'{\i}sica, Universidade de Bras\'{\i}lia, Bras\'{\i}lia-DF 70919-970, Brazil}
\author{Railson Vasconcelos}
\affiliation{Instituto de F\'{\i}sica, Universidade de Bras\'{\i}lia, Bras\'{\i}lia-DF 70919-970, Brazil}
\author{Fujun Liu}
\affiliation{Instituto de F\'{\i}sica, Universidade de Bras\'{\i}lia, Bras\'{\i}lia-DF 70919-970, Brazil}
\author{S-J Xie}
\affiliation{School of Physics, State Key Laboratory of Crystal Materials, Shandong University, Jinan 250100, China}
\author{Hao Zeng}
\email{haozeng@buffalo.edu}
\affiliation{Department of Physics, University at Buffalo, the State University of New York, Buffalo, NY 14260, USA}
\date{\today}

\maketitle
We developed a comprehensive theoretical framework focusing on many-body effects of exciton species in monolayer WS$_2$, including bright and dark excitons, and intra- and inter-valley biexcitons, to investigate valley dynamics in monolayer WS$_2$ subjected to a tilted magnetic field $\textbf{B}$. 
In particular, the evolution of the exciton population densities and the many-body particle scatterings are considered to calculate the valley polarization ($VP$) as a function of the magnetic field. 
We found valley splittings for the dark exciton and biexciton energy levels that are larger than those of bright excitons, of -0.23 meV/T. For example, -0.46 meV/T for dark excitons and -0.69 meV/T for bright-dark intra-valley biexcitons. 
Furthermore, inter-valley bright-dark excitons have an opposite valley energy splittings of +0.23 meV/T. For the samples pumped by linearly polarized light, $VP$ exhibits distinct magnetic field dependence for different types of many-body particle states. Among them, the $VP$ of the intra-valley bright-dark biexcitons increases with increasing magnetic field and reaches nearly 50$\%$ at B=65 T. The brightened dark exciton, on the other hand, exhibits vanishing $VP$, indicating long valley relaxation time. Remarkably, the inter-valley bright-dark biexciton shows unconventional behavior with an inverted $VP$. The opposite $VP$ for intra- and inter-valley bright-dark biexcitons, coupled with their large valley splitting and long valley lifetime may facilitate their coherent manipulation for quantum computing.

\section{Introduction}
\begin{center}
\begin{figure}
\includegraphics[width=\linewidth]{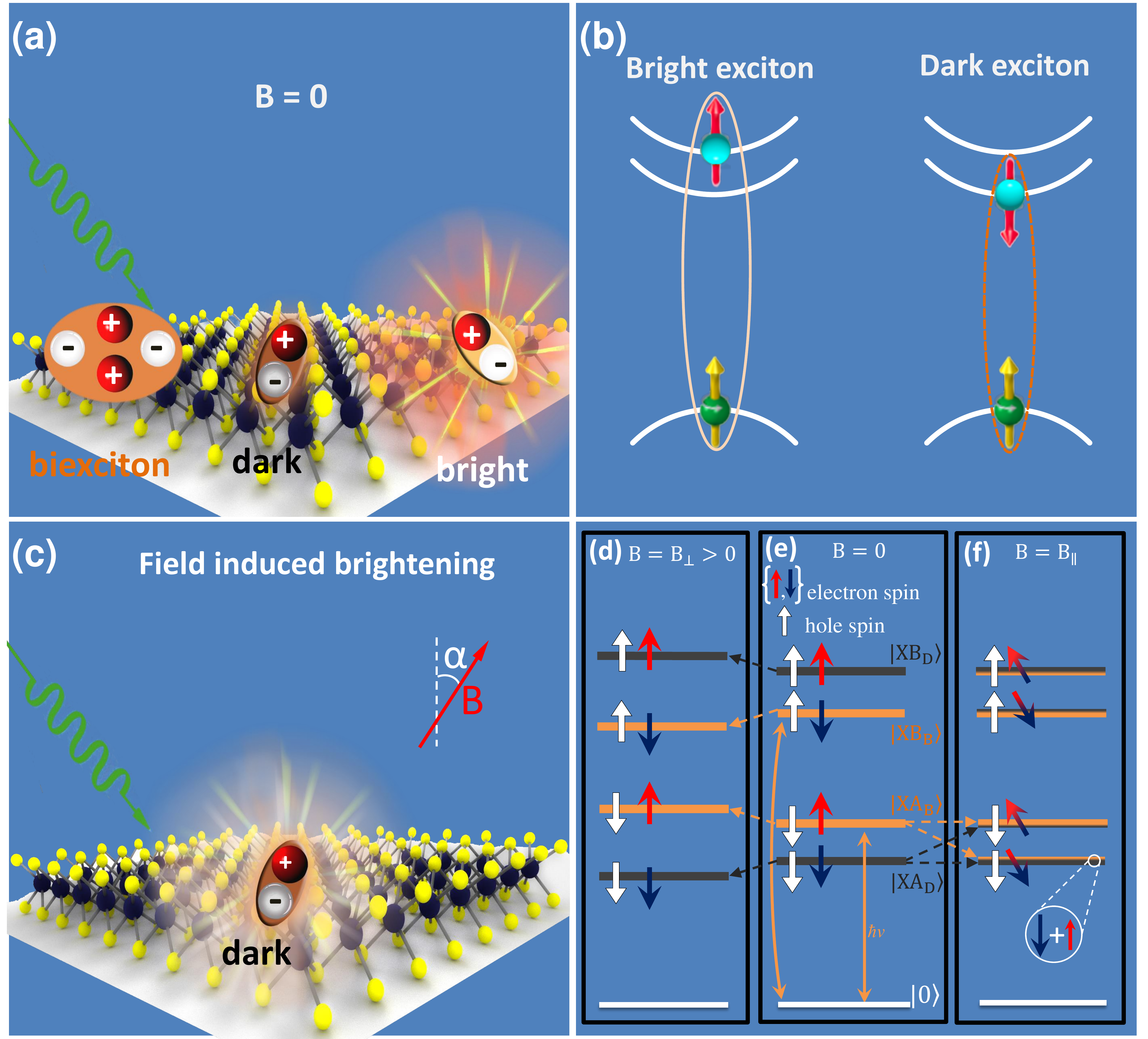}
\caption{Schematic representation of a monolayer WS$_2$ under an external magnetic field. (a) Coexistence of bright and dark excitons as well as biexcitons. (b) Electronic configurations of bright (left hand) and dark (right hand) A excitons. (c) DE brightening induced by  a tilted field $\textbf{B}$ with angle $\alpha$. 
(d)-(e) Schematic representation of the bright and dark exciton energy levels at the K valley.  Filled (blue and red) and empty arrows represent electron and hole spins, respectively, and orange arrows indicate the selection rule allowed optical transitions from the vacuum state ($|0\rangle$). (d) Shift of exciton energy levels caused by an out-of-plane magnetic field. (e) Zero field dark and bright states of A exition (lower part) and of B exciton (upper part). (e) Mixture of bright and dark states due to electron spin rotations driven by an in-plane  magnetic field. }
\label{AbstFig}
\end{figure}
\end{center}

In recent years, the coherent properties of bright excitons (BE) confined in semiconductor quantum dots have been extensively explored~\cite{kodriano2012complete}. BEs might be useful as qubits since their coherent states can be initialized, controlled, and read out using picosecond optical pulses. However, owing to their relatively short radiative lifetime due to their large oscillator strength, the use of the BEs as quantum information carriers remains elusive. Spin-forbidden dark excitons (DEs) (Fig.~\ref{AbstFig}(b)), on the other hand, can form long-lived matter qubits with long radiation and valley times~\cite{heindel2017accessing,poem2010dark}. However, their optical manipulation is challenging. Recently, monolayer transition metal dichalcogenides (TMDs) have received great interest, as their spin-locked valley indices provide an additional degree of freedom for quantum information processing~\cite{cao2012valley,wang2016control,jones2013optical,hao2017trion,hao2016direct,ye2017optical}. The small exciton's Bohr radius in TMD monolayers leads to a strong electron-hole exchange interaction~\cite{guo2018exchange} which will shorten the valley/spin relaxation time still further compared to that of excitons in conventional semiconductors quantum dots (of the order of sub-picoseconds). Thus, monolayer TMDs will face even greater challenges than quantum dots, if BEs were to be used as quantum information processing. The DEs, however, have vanishing electron-hole exchange interaction due to their anti-parallel spin configuration. The inter-valley scattering~\cite{zhang2017magnetic} of the DEs is thus highly suppressed. As a result, the DE has much longer radiative- and valley- lifetimes (tens of nanoseconds)~\cite{robert2017fine,jiang2018microsecond}, which makes it an appealing candidate for quantum information carrier. However, it is known that the DEs decouple from light, therefore, the challenge of optical read-out and control of the DE states has to be overcome before it can be readily used.

The strong Coulomb interaction in TMD monolayers make many-body effects more prominent than those in conventional semiconductors. This enables exploration of many-body particle states such as biexcitons, where two excitons are bound to a four-particle state~\cite{you2015observation,plechinger2015identification,sie2015intervalley,lee2016identifying, kim2016biexciton,okada2017observation,paradisanos2017room} (see Fig.~\ref{AbstFig}(a)). 
 Optically active intra-valley biexcitons can be composed of either two BEs (bright-bright intra-valley biexciton, XX) or one BE and one DE within the same valley, (bright-dark intra-valley biexciton, XX$_d$). Interestingly, aside from the intra-valley biexcitons, inter-valley bright-dark biexcitons XX$^{'}_d$ have also recently been observed experimentally in TMD monolayers~\cite{robert2018bright,chen2018coulomb,barbone2018charge,li2018revealing,ye2018efficient,nagler2018zeeman,hao2017neutral,pei2017excited} (see Fig.~\ref{Figschem} for a schematic representation of the different biexciton states).
To accurately describe the valley dynamics and quantitatively predict the valley polarization of exciton species in TMDs under a magnetic field, it is mandatory to consider the correlation among all these biexcitons as well as the single bright and dark excitons. Such studies are, however, still lacking, in spite of rapid growth of knowledge about these excitonic states.


It is known that out-of-plane ($B_{\perp}$)~\cite{aivazian:2015} and in-plane ($B_{\parallel}$) ~\cite{zhang:2017,molas2:2017} magnetic fields play different roles in excitonic emission and valley physics of monolayer TMDs. The former induces a valley energy splitting,
which is evaluated by the difference between the exciton energy level 
in K and K$^{\prime}$ valleys, i.e., $\bigtriangleup E^{KK^{\prime}}_i$=E$^K_{i}$-E$^{K^{\prime}}_{i}$, where E$^K_{i}$ (E$^{K^{\prime}}_{i}$) is the 
energy of the $i-th$ quasiparticle (BE, DE and biexcitons) in the K (K$^{\prime}$) valley ~\cite{stier2016exciton,zhang:2017,macneill2015breaking,srivastava2015valley,arora2016valley}. 
The valley splitting causes a charge transfer from one valley to the other.   Hence $B_{\perp}$ alters inter-valley scatterings of the exciton species tuning the $VP$, which can be measured experimentally by polarization-resolved magneto-photoluminescence (PL) and magneto-reflectance
spectroscopies~\cite{li2014valley,zou2018probing}. On the other hand, $B_{\parallel}$ mixes the bright and dark spin states ~\cite{donck:2018}, see Fig.~\ref{AbstFig}(f), leading to DEs brightening (see Fig.~\ref{AbstFig}). Thus light emission from these otherwise optically inaccessible states happens, and provides a pathway to optical read-out and control of the DE states ~\cite{donck:2018}.

In this work, we develop a comprehensive theoretical framework to investigate the valley physics, in particular valley splitting and valley polarization of excitons and biexcitons, in monolayer WS$_2$ subjected to a tilted magnetic field $\textbf{B}$. The conduction and valence bands involved in the bright/dark single and biexciton states have different spin and orbital angular momentum characters in different valleys, leading to different Zeeman splitting and population rebalancing of various exciton species in a magnetic field.  By a proper theoretical treatment of the correlations (scatterings) between different exciton species, $VP$ of these states can be accurately predicted. We focus our discussion on both intra- and inter-valley bright-dark biexcitons valley physics, while the behavior of other quasiparticles such as brightened dark excitons and intra-valley bright-bright biexcitons are presented as references for comparison purposes. These two types of biexcitons are highly relevant for quantum computing, as they possess the character of DE with orders of magnitude longer valley lifetime than that of BE, while at the same time allowing optical control of valley states, and thus are advantages over either BE or DE for information processing.   They show contrasting magnetic dependencies of the $VP$: namely, intravelley biexcitons exhibit a linearly increasing $VP$ with increasing magnetic field; while inter-valley bright-dark biexcitons show an unusually inverted $VP$. These interesting behaviors depend on  the critical role of dark excitons and correlations among exciton species. The coherent superposition of the two types of biexcitons may be used to construct valley pseudospin qubits with controllable phases.  
  

\section{Results}\label{SecResults}

The magneto-PL spectra, the DE brightening, the valley splitting and polarization of the exciton species including intra-valley BE, DE, XX and XX$_d$, and inter-valley XX$^{'}_d$ in a monolayer WS$_2$ under a titled magnetic field $\textbf{B}$ are obtained by using our comprehensive exciton valley dynamics theory. 
For details of the physical model, see Sec. \textbf{Methods} and the Supplementary Information.

\begin{figure} [h]
\begin{center}
\includegraphics[width=\linewidth] {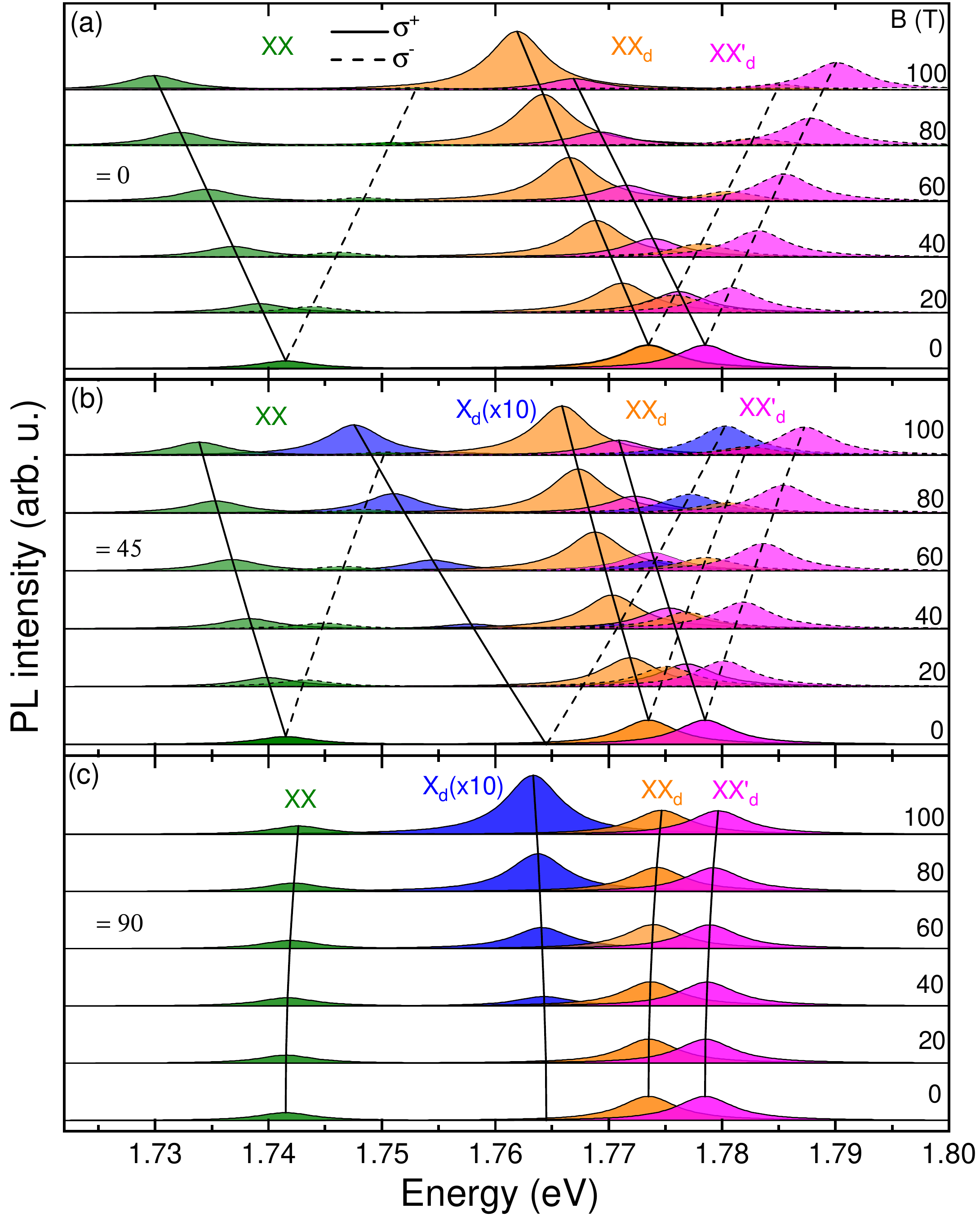}
\caption{\qu{Room temperature PL spectra of excitonic states in monolayer WS$_2$ excited by a linearly polarized light under magnetic fields in (a) Faraday  ($\alpha=0^o$) (b) tilted  ($\alpha=45^o$), and (c) Voigt ($\alpha=90^o$) geometries. X$_d$,  XX, XX$_d$, and XX$^{\prime}_d$ represent DE, bright-bright and bright-dark intra-valley biexcitons, and bright-dark inter-valley biexcitons respectively. Solid (dotted) lines correspond to emissions with $\sigma^+$ ($\sigma^-$) polarization. For clarity, the bright exciton emissions are omitted. The complete figure including bright excitons emission peaks can be found in the Supplementary Fig. 2.}}
\label{Spec1}
\end{center}
\end{figure}

%
%
%
%
Figure \ref{Spec1} illustrates the magnetic field dependence of the polarization-resolved PL spectra for monolayer WS$_2$ excited by a linearly polarized light under three different magnetic field directions.
In order to highlight our main results, only behaviors of the brightened DE and biexcitons are displayed in the figures of the main text. The complete PL spectra including the bright exciton emission peaks can be found in the Supplementary Fig. 2 (we also show, in Supplemnetary Fig. 3, a zoom around the $XX_d$ and $XX'_d$ emission peaks, in a case of narrower spectra width - obtained at lower temperature, for instance - in such a way that the two peaks can be clearly resolved).
All optically active excitonic emissions are polarization independent at zero field due to the valley degeneracy. However, this degeneracy is lifted by an out-of-plane ($\alpha=0^o$) magnetic field ($B_{\perp}$), as shown in Fig.~\ref{Spec1}-(a) in which the solid and dashed lines correspond to the emissions with $\sigma^+$ and $\sigma^-$ polarization, respectively. Since $\sigma^+$ ($\sigma^-$) couples exclusively to the K (K$^{\prime}$) valley, the emission indicated by the solid (dashed) line stems from the K (K$^{\prime}$) valley~\cite{yao2008valley}. 
Note that the energy difference between spectral peaks stemmed from K and K$^{\prime}$ valleys depends on the valley splitting of both the initial and the final states involved in the optical transition (see the vertical lines of Fig.~\ref{Figschem}, panels (b), (d), and (f)).
Furthermore, the height of the peaks for $\sigma^+$ is also different from that of the $\sigma^-$, which indicates a magnetic field induced valley  polarization.

Since $B_{\parallel}$ mixes the two spin
states in the conduction band (see methods section for a detailed discussion), the spin projections of the conduction and valence bands forming a DE are no longer perfectly antiparallel. The exciton dipole matrix element between these two bands then becomes finite and the dark excitons become brightened. Figure~\ref{Spec1}-(c) illustrates the magnetic field  dependence of exciton emissions in a monolayer WS$_2$ subjected to an in-plane magnetic field ($\alpha=90^o$) at room temperature. 
As the magnetic field increases, a new emission peak at 1.76 eV originated from the dark exciton $X_d$ emerges. A similar peak was also detected in carbon nanotubes through magneto-PL spectroscopy~\cite{srivastava2008direct}. 
With increasing field magnitude, the intensity of the brightened DEs increases.
Also notice that the in-plane magnetic field induces neither peak shifts nor valley splitting.



To reveal the combined effects of in-plane and out-of-plane field components, Fig.~\ref{Spec1}-(b) shows the magnetic field dependence of the PL spectra of WS$_2$ under a tilted magnetic field with $\alpha=45^o$. As expected, at a finite magnetic field, the DE brightening and the breaking of the valley degeneracy take place simultaneously. 
Also notice that the PL spectra are determined by the interplay between in-plane and out-of-plane magnetic field components. Dark exciton brightening, for instance, is primarily a consequence of the in-plane field component. 
The out-of-plane component, however, changes the bright-dark energy separation in K (K$^{\prime}$) valley, $\Delta E_{BD}^{K(K^{\prime})}=E^{K(K^{\prime})}_{BE}-E^{K(K^{\prime})}_{DE}$, and thus influences the spin mixing in the conduction band and thus the DE brightening. 

%



\begin{figure} [h]
\begin{center}
\includegraphics[width=0.95\linewidth] {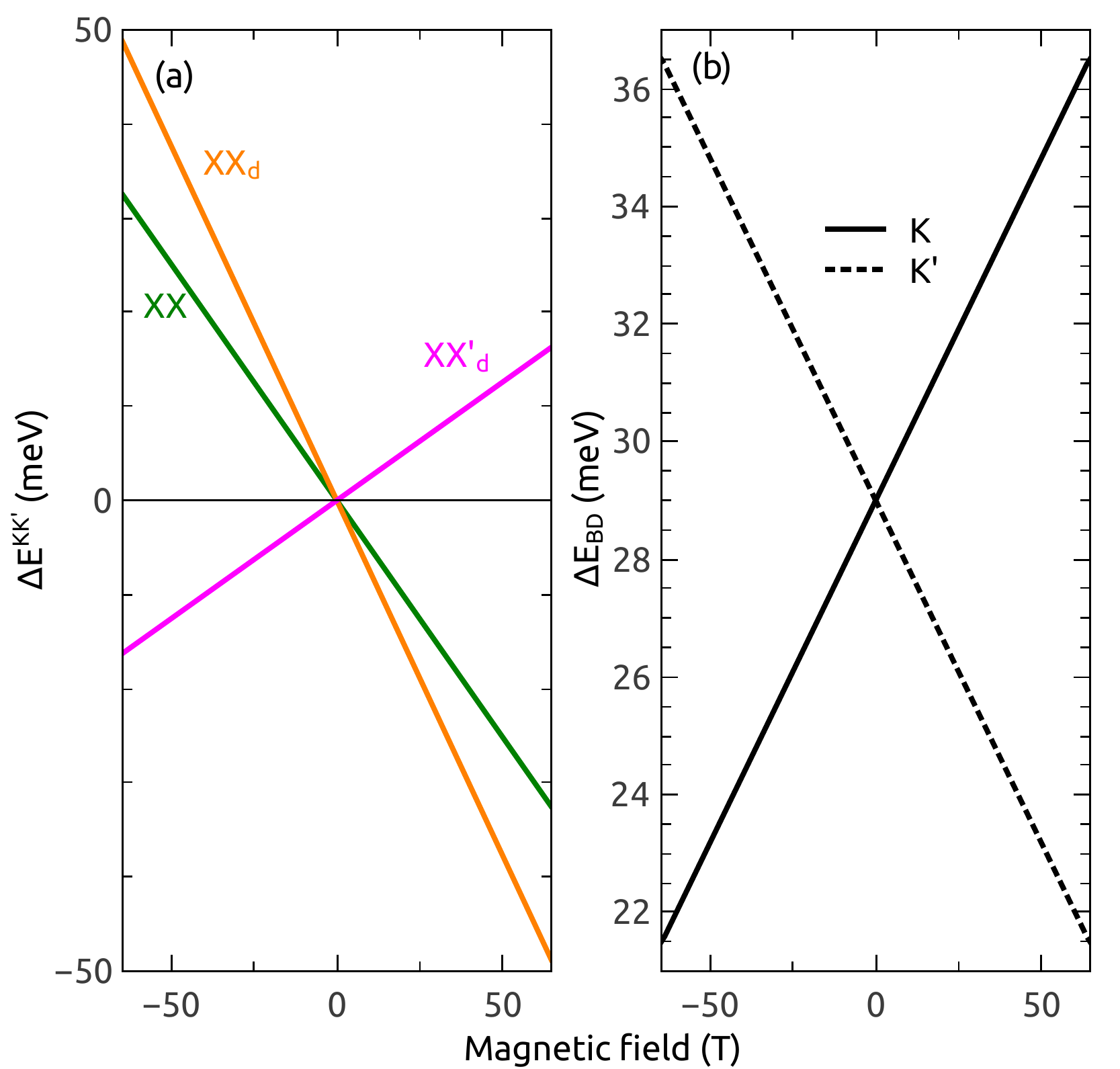}
\caption{Effects of an out-of-plane magnetic field to the exciton valley dynamics in a monolayer WS$_2$. (a) Valley  splittings, $\Delta E^{KK'}$, of intra-valley (XX and XX$_d$) and inter-valley (XX$^{\prime}_d$) biexcitons (the horizontal $VP=0$ line serve as a guide to the eyes) and (b) bright-dark energy separation ($\Delta E_{BD}$) in K (solid line) and K$^{\prime}$ (dotted line) valleys as a function of an out-of-plane magnetic field.}
\label{FigValleySplit}
\end{center}
\end{figure}

In order to gain quantitative insights into these results, Fig.\ref{FigValleySplit} (a) shows the valley splitting $\bigtriangleup E^{KK^{\prime}}$ of the biexciton states as a function of the out-of-plane field. It is evaluated by the difference between the total energy of corresponding excitonic quasiparticle states in K and K$^{\prime}$ valley, which, for biexcitons, is different from the spectrum splitting shown in Fig~\ref{Spec1}~\cite{nagler2018zeeman}.
The Zeeman energy shift comprises the contributions from the spin and orbital magnetic moment~\cite{aivazian:2015} (black and orange lines in Fig.~\ref{Figschem}, respectively). The quantum numbers of spin and orbital angular momenta are inverted in the two valleys, therefore the biexciton energy exhibits valley-dependent shifts. 
Notice that different excitonic quasiparicles exhibit distinct splittings. The bright-dark intra-valley biexciton exhibts the larger splitting (-0.69 meV/T), while its inter-valley conterpart shows an inverted dependence on the external field (+0.23 meV/T).
The difference can be understood through the analysis of the single exciton splitting; the biexciton energy is given by the sum of the two constituent exciton energies minus the field independent biexciton binding energy.  
When an out-of-plane field is applied, the conduction and valence band energies shift due to the Zeeman effect of an electron spin and orbital angular momentum, as shown in Fig.~\ref{Figschem}.
For bright excitons, the shift induced by the spin has the same value for the conduction and valence bands, so it does not contribute to the net change in the optical transition energies.
Thus, only the orbital magnetic moment of the valence bands, which differs for the two valleys, producing a measurable shift. 
Specifically, the valence band has orbital angular moment $m=2$ and $-2$ in K and K$^{\prime}$ valleys, respectively, then the orbital angular momentum contribution leads to a valley splitting of -0.23 meV/T for bright excitons~\cite{koperski2018orbital,aivazian:2015}.
For DEs, however, the involved valence and conduction band states have opposite spins. Both the spin- and the orbital-induced band shifts contribute to changes in the optical transition energy. As a result the net shift of DEs (-0.46 meV/T) is twice as large as that of BE.
The valley splitting of the XX state is also twice as large as the splitting of the BE state (-0.46 meV), while the valley splitting of the intra-valley bright-dark biexcitons is the largest, with $\Delta E^{KK^{\prime}}_{XX_d}=-0.69$ meV/T. Figure~\ref{Figschem}(a)-(d) shows a schematic representation of the band energy shifts (see panels (a) and (c)) and the intrvalley biexcitons and single excitons valley splittings (panels (b) and (d)). Finally, the valley splitting for the inter-valley bright-dark exciton is $\Delta E^{KK^{\prime}}_{XX'_d}=+0.23$ meV/T, which shows an inverted sign. Such a valley splitting is in accordance with the total $g$ factor, $g_T=\Delta E^{KK^{\prime}}_{XX'_d}/\mu_B \sim 4$, of bright-dark biexcitons in WS$_2$ monolayer determined by PL measures~\cite{nagler2018zeeman}. A schematic representation of the band Zeeman shift and the valley splitting of inter-valley biexciton is shown in Fig.~\ref{Figschem}(e)-(f). 

\begin{figure} [h]
\begin{center}
\includegraphics[width=\linewidth] {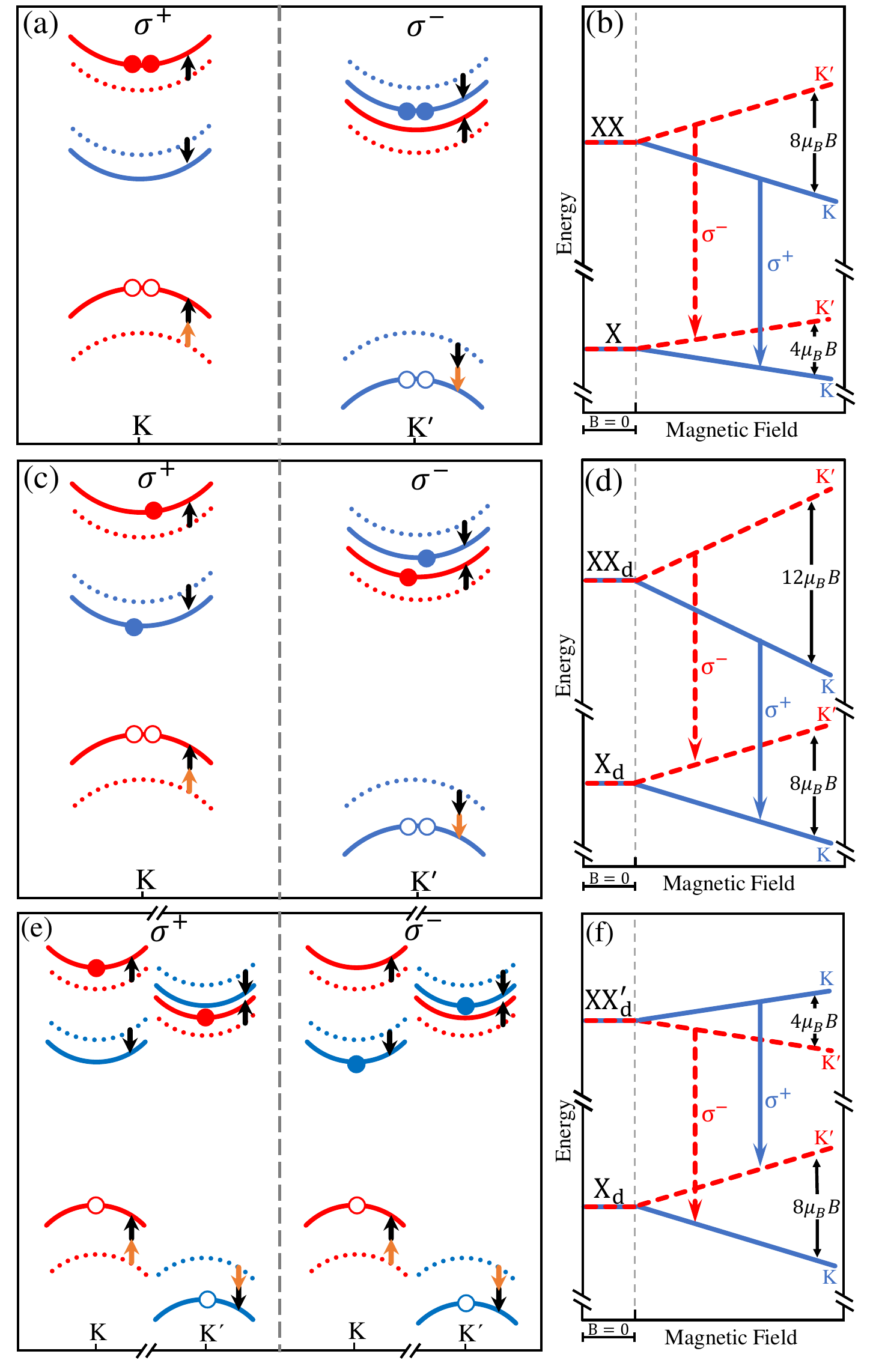}
\caption{Schematic representation of the biexciton fine structure. \qu{Left panels: Electronic configurations of intra-valley (a) bright-bright (XX) and (c) bright-dark (XX$_d$) biexcitons as well as (e) inter-valley bright-dark biexcitons (XX$^{\prime}_d$) in the absence (dotted lines) and presence (solid lines) of magnetic field in K (left hand) and K$^{\prime}$ (right hand) valleys. Filled and empty circles represent electron and hole, respectively. Black and orange arrows indicate magnetic moment of spin and orbital angular momentum. 
Because we are interested in A-exciton species, only the lowest valence band in each valley is shown.
Right panels: Energy levels of excitons and biexcitons in K (solid lines) and K$^{\prime}$ (dashed lines) valleys in the zero and finite magnetic field. The red and blue vertical arrows indicate optical transitions. 
}}
\label{Figschem}
\end{center}
\end{figure}

%

Aside from the valley splitting, the intra-valley bright-dark energy separation, $\Delta E_{BD}$, can also be tuned by an out-of-plane magnetic field, as shown in Fig.~\ref{FigValleySplit}-(b). $\Delta E_{BD}$ 
determines the electronic configuration of excited states and the scatterings among them, through the Boltzmann distribution balancing exciton populations between bright and dark excitons, with $k_b$ the Boltzmann constant and $T$ the temperature (see the Supplementary Information for more details).
Both the magnetic field induced valley splitting and the magnetic-tunable bright-dark energy separation have important effects on the exciton dynamics, as will be described in the following.

\begin{figure} [h]
\begin{center}
\includegraphics[width=\linewidth] {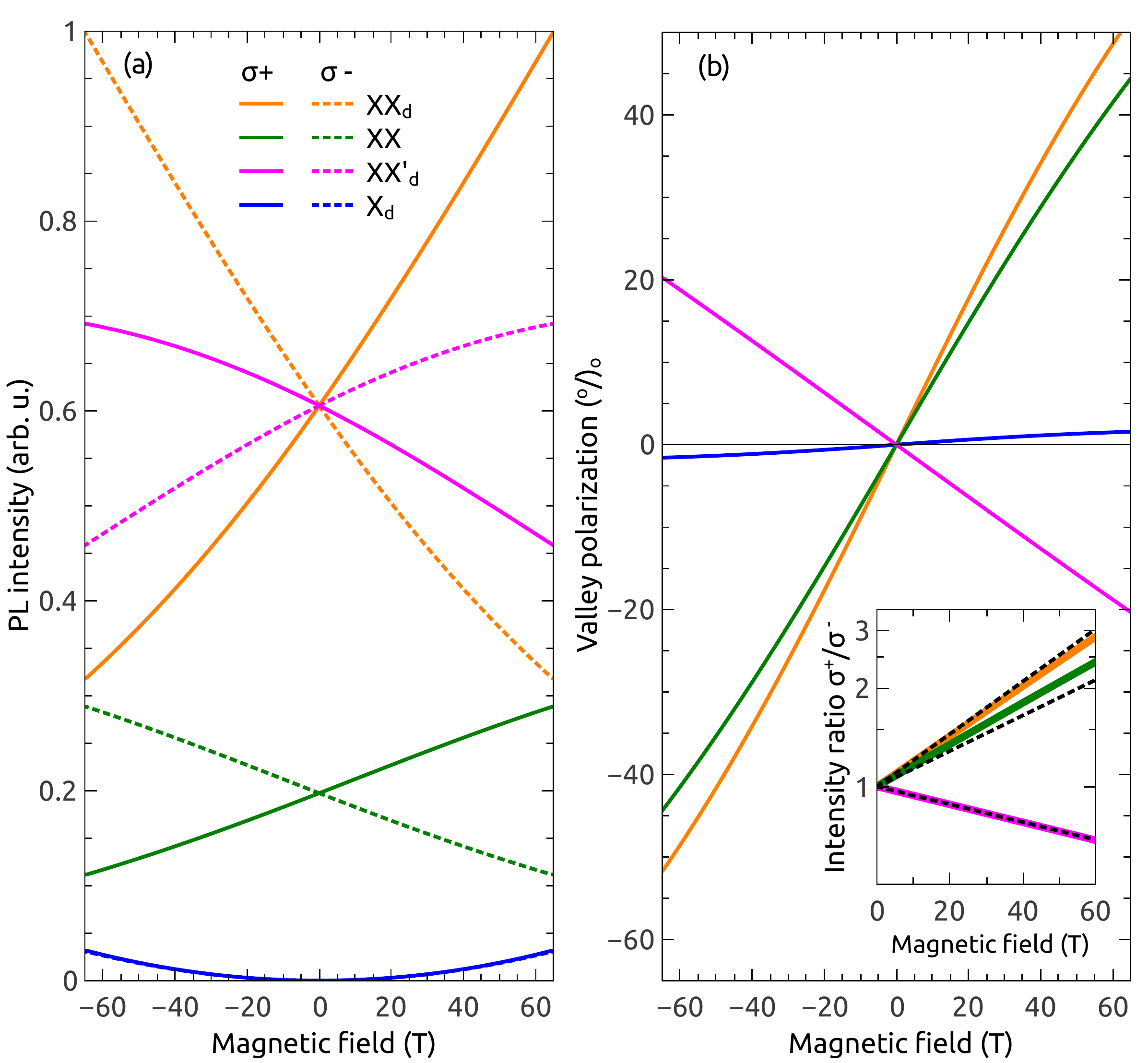}
\caption{Magnetic-PL intensity and valley polarization as a function of a tilted magnetic field. (a) Room temperature PL intensities of X$_d$, XX, XX$_d$, and XX${\prime}_d$ excitons and biexcitons in a monolayer WS$_2$ as a function of magnetic field with a tilted angle $\alpha=45^o$. Solid (dotted) line corresponds to PL intensity of polarized emission with $\sigma^+$ ($\sigma^-$ ) helicity, related to optical transitions in K (K$^{\prime}$) valley states. (b) The corresponding valley polarization degree. The horizontal $VP=0$ line serve as a guide to the eyes. Inset: Corresponding intensity ratio (log. scale) of the $\sigma^+$ and $\sigma^-$ polarized emission. The dotted lines represent the thermal Boltzmann distribution, $exp(-\Delta E^{KK^{\prime}}B_{\perp}/k_bT)$, at $T=300$ K, with $\Delta E^{KK^{\prime}}$ the valley splitting per tesla.}
\label{FigPLint}
\end{center}
\end{figure}

To further explore the effect of the interplay between $B_{\parallel}$ and $B_{\perp}$ on dark exciton brightening and biexciton emission, we investigate the polarization-resolved PL intensity as a function of a tilted magnetic field ($\alpha=45^o$), for the monolayer WS$_2$ excited with a linearly polarized light. 
In Fig.~\ref{FigPLint}(a) the solid (dashed) lines represent the PL intensities of $\sigma^+$ ($\sigma^-$) polarized emission of the exciton species. 
At a finite field, the brightened DE emission emerges in both valleys and increases parabolically with the field (see the blue line in Fig.~\ref{FigPLint}). 
In contrast, the PL intensity of the bright-bright biexciton state XX exhibits opposite magnetic field dependence in K and K$^{\prime}$ valleys (green solid and dotted lines in Fig.~\ref{FigPLint}, respectively), featuring a ``X'' pattern crossing at $B=0$. This is consistent with experimental observation~\cite{aivazian:2015} and theoretical prediction~\cite{vasconcelos2018dark}. As the valley splitting increases with increasing field, the PL intensity stemmed from the lower energy level increases while the one from the higher energy level decreases. The PL intensity of bright-dark biexcitons (XX$_d$ and XX$^{\prime}_d$) under a  tilted magnetic field exhibits a ``bent-X'' pattern. Interestingly, intra-valley and inter-valley bright-dark biexcitons display opposite slopes, that is, for $B> 0$, the intensity increases in the K valley for bright-dark intra-valley biexcitons and decreases for bright-dark inter-valley states (the valley that labels the inter-valley biexciton is the one that holds the BE).
It is interesting to note also that, at B=0, the emission of the bright-dark biexcitons is around three times more intense than that of the XX. It is ascribed to the longer lifetime of dark excitons, which leads to a higher probability of formation of XX$_d$ and XX$^{\prime}_d$ states than XX species. Furthermore, the bright-dark intravelley biexciton shows a stronger dependence on the magnetic field. To interpret this behavior we need to consider the occupation number of both the initial (XX$_d$) and final (X$_d$) states in the entire optical emission process. For $B>0$, the number of the dark excitons in the K-valley is reduced because of the larger bright-dark exciton energy level separation (see Fig.~\ref{FigValleySplit}(b)). As the magnetic field increases, this population imbalance also increases. Thus an increase in the transition probability from the initial XX$_d$ state to the final X$_d$ in the K-valley is expected. As a result, an enhancement in the PL intensity of the XX$_d$ with increasing field in the K-valley is observed. 
The field dependence of PL intensity can be used to determine the nature of biexcitons.

Realizing valley polarization (selective occupation of a particular valley) is the important first step for manipulating the valley degree of freedom for information processing.
The degree of valley polarization is proportional to the difference in PL intensities between $\sigma^+$ and $\sigma^-$ emissions, and is defined by $VP_{j}=\frac{I_{j}(K)-I_{j}(K')}{I_{j}(K)+I_{j}(K')}$, where $I_{j}(K^{(')})$ is the PL intensity of the $j-th$ exciton specie in the K (K$^{\prime}$) valley.
In the presence of a magnetic field, we assume that the inter-valley scatterings are composed of two consecutive process: a spin-flip process followed by an energy relaxation, emitting a phonon~\cite{jiang2017zeeman}. The reverse process must first involve a phonon absorption followed by a spin-flip (see the Supplementary Information Sec. I for more details). 
Since the phonon absorption is suppressed by the Boltzmann factor, $\exp(-\Delta E^{KK'}/ k_bT)$, for a valley splitting of $\Delta E^{KK'}$, 
the intensity of excitonic quasiparticle with lower energy is dominant. This effect can be seen in Fig.~\ref{FigPLint}(b) that displays room temperature $VP$ of brightened DEs and three types of biexcitons in a monolayer WS$_2$ as a function of tilted ($\alpha=45^o$) magnetic field. For the system pumped with a linear polarized light, the $VP$ of all the quasiparticles (X$_d$, XX$_d$, XX and XX$^{'}_d$) exhibit a near-linear dependence on the external magnetic field. However, for different quasiparticles, the $VP$ behaviors are distinctively different. 
For intra-valley biexcitons (both XX$_d$ and XX), the $VP$ increases linearly with increasing magnetic field, reaching around 50$\%$  at B=65 T for XX$_d$. This is attributed to negative valley splitting of the BE, DE and intra-valley biexcitons (XX and XX$_d$) induced by a positive magnetic field, as shown in Fig.~\ref{FigValleySplit}(a).  The lower energy of these states in K valley than that in K$^{\prime}$ valley favors a charge transfer from the
K$^{\prime}$ valley to the K valley. Consequently, the $VP$ becomes magnetically enhanced. Surprisingly, the X$_d$ posses a tiny $VP$ (1.4$\%$ at B=60 T), despite a large VS. This is attributed to its long inter-valley scattering time~\cite{jiang2018microsecond}, which keeps the DE in its initial valley, i.e., little charge transfer between the valleys occurs. Therefore, the X$_d$ possesses similar emission intensity in both valleys.  The inter-valley bright-dark biexciton XX$^{\prime}_d$, however, shows an unexpected inverted valley polarization, i.e. $VP$ is negative at positive fields. This behavior is explained by the total energy of the biexciton, which determines the population distribution over the valley split states. Accounting for both bright and dark exciton energies which contribute to the biexciton total energy, one finds a positive valley splitting (for $B>0$), that is, the 
the XX$^{\prime}_d$ state (BE in the K valley and DE in the K$^{\prime}$ valley, $\sigma^+$ emission) has higher energy than the
X$^{\prime}$X$_d$ state (BE in the K$^{\prime}$ valley and DE in the K valley, $\sigma^-$ emission). Therefore, the latter is more populated, leading to an inverted polarization. In addition, the intensity ratio $I_{\sigma^+}/I_{\sigma^-}$ follows the room temperature Boltzmann distribution $exp(-\Delta E^{KK^{\prime}} B_{\perp}/k_bT)$ [see the inset in Fig.~\ref{FigPLint}(b)]. It indicates that the thermal redistribution of biexcitons in K and K$^{\prime}$ as well as the valley polarization is mainly determined by the valley splitting. It is supported by polarization-resolved PL spectroscopy measurements~\cite{nagler2018zeeman} (see Supplementary Fig. 4 for a direct comparison between our theoretical results and experimental data).

%


\begin{figure*} [t]
\begin{center}
\includegraphics[width=\linewidth] {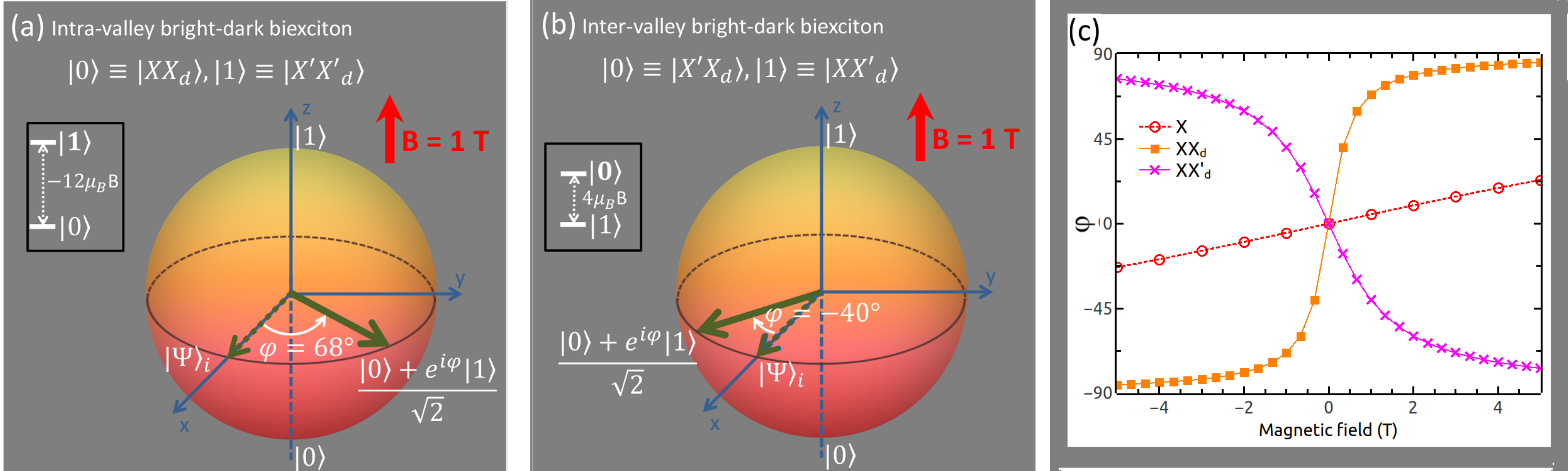}
 \caption{Schematic representation of the valley pseudospin manipulation via an external magnetic field. (a)-(b) Bloch sphere representing a qubit composed of two biexcitons with opposite circularly polarized light. At B=0, a linearly polarized optical pulse excites a coherent superposition of biexcitons in K and K$^{\prime}$ valleys. (a) Negative valley splitting of intra-valley bright-dark biexcitons induced by an out-of-plane magnetic field and a clockwise rotation of the valley pseudospin in the equatorial plane of the Bloch sphere. (b) Positive valley splitting of inter-valley bright-dark biexcitons induced by an out-of-plane magnetic field and an anticlockwise rotation of the valley pseudospin in the equatorial plane of the Bloch sphere.
 (c) Rotation angle of the valley pseudospin as a function of the magnetic field for XX$_d$ and XX$^{\prime}_d$ exciton species. We also include results for the X state for comparison purposes.  }
\label{Bloch}
\end{center}
\end{figure*}

A tilted magnetic field induces both brightening of dark excitons and its large valley splitting in monolayer WS$_2$. Such dark exciton states can be attractive as quantum information carriers due to their lifetime orders of magnitude longer than that of Bright excitons. The magnetic field brightening makes them optically accessible and controllable. A linearly polarized light will be able to initialize a valley pseudospin qubit, which is a coherent superposition of valley pseudospins up and down. 
An out-of-plane magnetic field can rotate the pseudospin with the angle of rotation determined by valley splitting. A larger valley splitting together with a longer valley coherence time means higher angle of rotation for the pseudospin on the Bloch sphere. Furthermore, the ultrastrong Coulomb interactions in TMDs make many-body effects such as biexcitons prominent. Most notably, the bright-dark biexcitons allow one to take advantage of both the long valley lifetime and large valley splitting for pseudospin manipulation, and strong light-matter interaction for optical initialization and readout. The intra- and inter-valley bright-dark biecitons exhibit opposite valley splitting, which means the exciton pseudospins rotate in opposite directions when subjected to a magnetic field. A schematic representation of qubits constructed from the intra- and inter-valley bright-dark excitons on the Bloch sphere and the dependence of the rotation angles of the pseudospins on the external perpendicular magnetic field can be seen in Fig.~\ref{Bloch}. 
We define pseudospin $|0\rangle \equiv |XX_d\rangle$ and pseudospin $|1\rangle \equiv |X^{\prime}X^{\prime}_d\rangle$ for the intra-valley bright-dark biexciton. For the inter-valley counterpart, $|0\rangle \equiv |XX^{\prime}_d\rangle$ and $|1\rangle \equiv |X^{\prime}X_d\rangle$.  Assuming excited by a linearly polarized light, the initial state is a coherent superposition of $|0\rangle$ and $|1\rangle$, pointing along the x-direction. Applying a DC magnetic field causes a rotation of the pseudospin about the z-axis. The angle of rotation
$\varphi=$arctan$\left(\Omega T^* \right)$~\cite{wang2016control}, where the precession angular frequency is proportional to the valley splitting, given by $\Omega=\Delta E^{KK^{\prime}}/\hbar$. 
$T^*$ was reported to be 0.26~ps for bright excitons~\cite{schmidt2016magnetic}. Due to the presence of dark excitons, the valley coherence time for bright-dark bixcitons is expected to be orders of magnitude larger. Here we take $T^*$ to be 3.7~ps as a conservative lower bound estimate for calculating the precession angle. 
As can be see from Fig.~\ref{Bloch}(c), both biexcitons rotate by larger angles than that of the bright exciton, with the intra-valley biexciton possessing the largest angle of rotation. This is a consequence of both of its larger valley splitting and longer valley coherence time.   
One can imagine that using such excitonic states in coupled TMD quantum dots, it is possible to construct entangled states with controllable phases. 
The manipulability of the pseudospins can be further increased by using magnetic proximity effect~\cite{zou2018probing,zhao2017enhanced}.  
The valley splitting can be enhanced by  more than an order of magnitude, allowing the same angle of rotation to be achieved at fields much smaller than 1~T. This field can be readily supplied by a miniaturized inductive write head, allowing local control of individual quantum device.  
Therefore, our results suggest that bright-dark biexcitons in TMDs are promising candidates for quantum information applications. Such a scenario can motivate further theoretical and experimental effort in looking for protocols to deterministic initialization and read out of the valley pseudospins.

\section{Methods}

Polarization-resolved magneto-PL measurements have shown that an applied perpendicular magnetic field $B_{\perp}$ couples with the spin and orbital magnetic momenta and leads to a Zeeman shift of the single-particle band edges~\cite{aivazian:2015}. Among them, the shift due to the spin magnetic moment is given by $\Delta_s = g_s s_z \mu_BB_{\perp} $, with  $\mu_B$ the  Bohr magneton, $g_s=2$ the spin $g-$factor, and $s_z=\pm 1/2$ the electronic spin. The orbital Zeeman shift is mainly due to the tungsten $d$-orbitals, $\Delta_a = m_l\mu_BB_{\perp}$, where
$m_l$ is the magnetic quantum number. In the first order approximation, it does not affect the conduction bands, since $m_l \sim 0$. It shifts, however, the valence bands, which primarily consist of $d$-orbitals with $m_l=2$ in the $K$ valley and $m_l=-2$ in the $K^\prime$ valley, respectively. The exciton energy is then given by the difference between field dependent conduction and valence band energies, subtracted by the field independent binding energy~\cite{scharf2017magnetic} [Fig~\ref{AbstFig}(d)-(e)].
The energy of bright, $E_{X^{\tau}}$, and dark, $E_{X_d^{\tau}}$, states at $\tau$ valley are given by

 \begin{eqnarray}
  E_{X^{\tau}}&=&Eg-Eb_{XA}+\frac{\lambda_c}{2}-\frac{\lambda_v}{2}-\tau \Delta_a\nonumber \\
  &=& Eg-Eb_{XA}+\frac{\lambda_c}{2}-\frac{\lambda_v}{2}-\tau 2\mu_BB_{\perp}
\end{eqnarray}
and
\begin{eqnarray}
  E_{X_d^{\tau}}&=&Eg-Eb_{XA}-\frac{\lambda_c}{2}-\frac{\lambda_v}{2}-\tau 2\Delta_s -\tau \Delta_a \nonumber \\
  &=& Eg-Eb_{XA}-\frac{\lambda_c}{2}-\frac{\lambda_v}{2} -\tau4 \mu_B B_{\perp},
 \end{eqnarray}
where $Eg$ represents the band gap, $\lambda_{c(v)}$ corresponds to the energy splitting in the conduction (valence) band due to spin-orbit interaction, $E_{b\Gamma}$ is the binding energy of the $\Gamma=\{X, X_d\}$ state, and $\tau$ is the valley index ($\tau=1$ for K and $\tau=-1$ for K$^{\prime}$ valley).

 The biexciton energy is evaluated by the total energy of two constituent excitons minus its binding energy. For intra-valley bright-bright, bright-dark, and inter-valley bright-dark biexcitons, they are given by
  \begin{center}
     \begin{eqnarray}
     E_{X^{\tau}X^{\tau}}&=& 2E_{X^{\tau}}-Eb_{XX},
    \nonumber \\
         E_{X^{\tau}X_d^{\tau}}&=& E_{X^{\tau}}+ E_{X_d^{\tau}}-Eb_{XX_d},
    \nonumber \\
    E_{X^{\tau}X^{\tilde{\tau}}_d}&=& E_{X^{\tau}}+ E_{X_d^{\tilde{\tau}}}-Eb_{XX'_d}. \label{EqBiexc},
   \end{eqnarray}
  \end{center}

It is also known that an external in-plane field $B_{\parallel}$ applied to TMD monolayers mixes the two spin states
in the lowest conduction band. This effect was observed experimentally through the brightening of initially dark (spin-forbidden) intra-valley excitonic states~\cite{zhang:2017}.

In a low-energy model, we can describe the effect of the in-plane field on the conduction band via perturbation theory as

\begin{eqnarray}
\hat{H}_{c}^{\tau}  &=& \hat{H}_{0,c}^{\tau} + \hat{V} \nonumber \\ &=&\begin{pmatrix}
 \frac{Eg}{2}+ \tau \frac{\lambda_{c}}{2}&0\\
0 &   \frac{ Eg}{2}- \tau \frac{\lambda_{c}}{2}
\end{pmatrix}
\nonumber \\ &+&
\begin{pmatrix}
0 & -\mu_B g_s (B_x - i B_y)\\
-\mu_B g_s (B_x + i B_y) & 0
\end{pmatrix},
\end{eqnarray}
where the matrix of single particle Hamiltonian is constructed based on the single-particle basis $\{|\psi_{c, \uparrow} \rangle, |\psi_{c, \downarrow}\}$. Upon diagonalizing $\hat{H}_{c}^{\tau}$, the energies are obtained by

\begin{eqnarray}
E_{\psi_{c, \uparrow}}^{\tau, mix}&=&  \frac{Eg}{2} + \tau \frac{\lambda_{c}}{2} + \tau \frac{g^2 \mu_B^2 |B_{\parallel}|^2}{4 \lambda_{c}}, \nonumber \\
E_{\psi_{{c}, \downarrow}}^{\tau, mix}&=& \frac{ Eg}{2} - \tau \frac{\lambda_{c}}{2}  - \tau \frac{g^2 \mu_B^2 |B_{\parallel}|^2}{4 \lambda_{c}}.
\label{ElectronEnergy} \end{eqnarray}
It is worthy to point out that the perturbative correction to the band energy due to the presence of in-plane magentic field is vanishingly small, even for large magnetic fields. The major effect of the $B_{\parallel}$ is to mix the different spin states, which can be seen by the new eigenstates:

\begin{eqnarray}
|\psi_{c, \uparrow}^{\tau, mix}\rangle &=& c_{\uparrow \uparrow}|\psi^{\tau}_{c, \uparrow}\rangle + c_{\uparrow \downarrow} e^{i \phi }|\psi^{\tau}_{{c}, \downarrow}\rangle, \nonumber \\
|\psi_{{c}, \downarrow}^{\tau, mix}\rangle &=& - c_{\downarrow \uparrow} e^{-i \phi }|\psi^{\tau}_{c, \uparrow}\rangle + c_{\downarrow \downarrow} |\psi^{\tau}_{{c}, \downarrow}\rangle,
\label{ElectronEigenvector} \end{eqnarray}
where $\phi$ is the azimuth angle of the magnetic field, defined by $\tan \phi=B_y/B_x$. $|c_{\uparrow \uparrow}|=|c_{\downarrow \downarrow}|= {1}/{\sqrt{1+W_c^2}}$ and $|c_{\uparrow \downarrow}|=|c_{\downarrow \uparrow}|= {W_c}/{\sqrt{1+W_c^2}}$ denote the spin conserving and spin-flip coefficients, respectively. Both of them depend on $B_{\parallel}$ through $W_c={g_s \mu_B B_{\parallel}}/{2 \lambda_c}$.

An analogous description can be made for the valence band, but the spin mixing is negligible due to very large spin-orbit splitting, i.e., $\lambda_v \gg \lambda_c \rightarrow W_v \ll W_c$. Hence, we neglect the spin mixing in the valence band.

In summary, an in-plane magneic field mixes spin-up and spin-down states in the conduction band, hybridizing the spin states. Consequently, the initially spin-forbidden electric dipole transition such as dark exciton emission becomes allowed with a probability proportional to $|c_{\downarrow \uparrow}|^2$. 

\begin{figure} [h]
\begin{center}
\includegraphics[width=\linewidth] {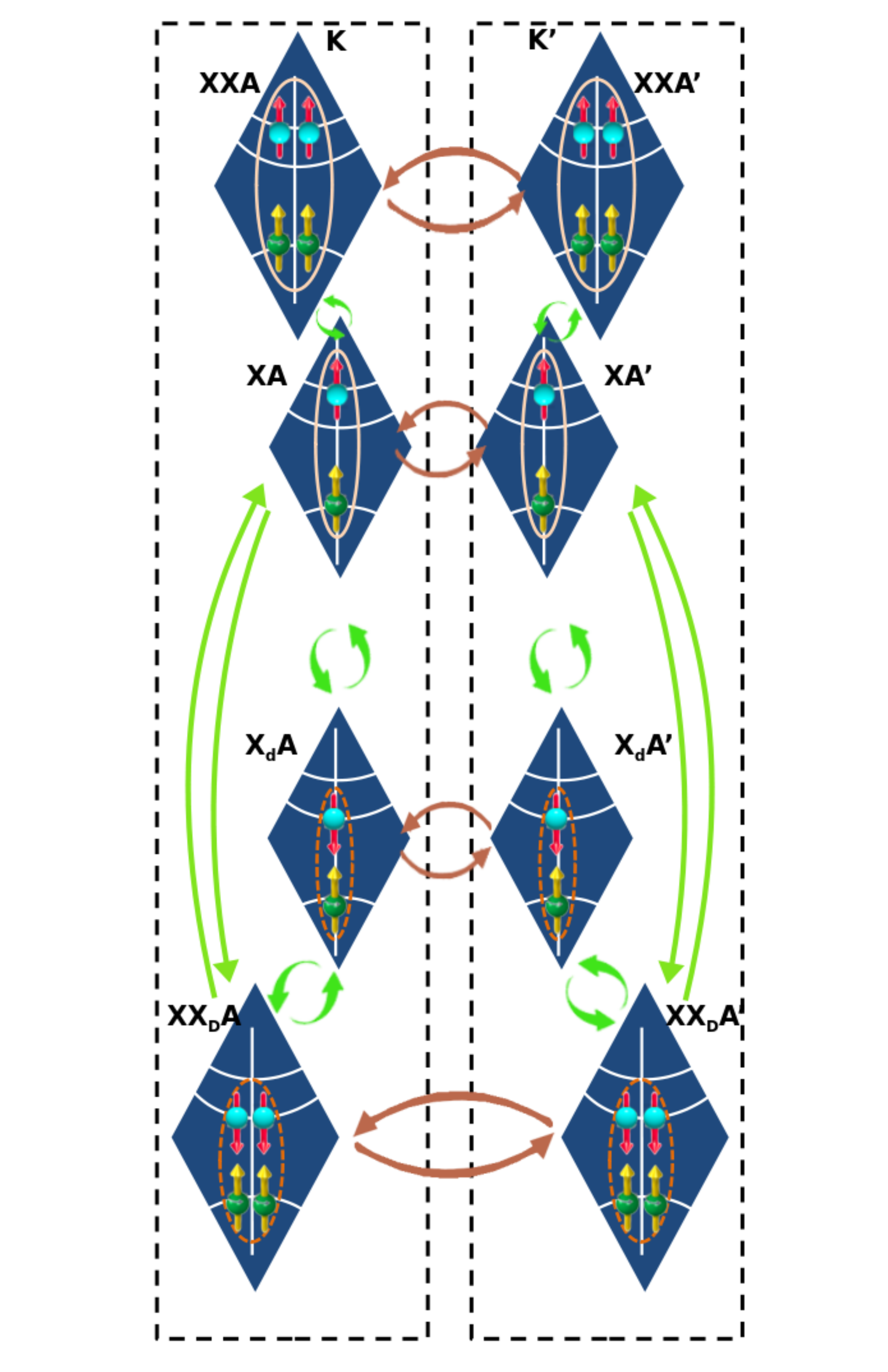}
\caption{Schematic representation of intra-valley (green arrows) and inter-valley scatterings (brown arrows). The left and right panels represent K and K$^{\prime}$ valleys, respectively. The blue and green circles denote electrons and holes, respectively. The vertical arrows indicate spin states. }
\label{Figsingleparticle}
\end{center}
\end{figure}

With the knowledge of the effect of the external magnetic field to the exciton energy levels, we propose a theoretical framework to describe the intra-valley and inter-valley involved valley dynamics. 
An schematic representation of the exciton species considered and the intra- and inter-valley scatterings between them is shown in Fig.~\ref{Figsingleparticle}. For clarity purpose, the inter-valley XX$^{\prime}_d$ biexciton and the involved scatterings are omitted in the schematic diagrams, although they are considered in our calculation.
Our comprehensive set of coupled rate equations takes into account (i) photon-generation of intra-valley bright excitons, (ii) radiative recombination of bright and brightened-dark excitons and biexcitons, (iii) exciton-to-biexciton transitions, (iv) bright-dark intra-valley scattering, via a phonon-mediated spin-flip process, and (v) inter-valley scattering of either excitons or biexcitons. The inter-valley scattering in the presence of the external magnetic field is described by a two-step process composed of a resonant spin-flip process and a phonon-mediated relaxation/excitation~\cite{jiang2017zeeman}, as mentioned in the previous section.  
Furthermore, DE brightening process is incorporated in our rate equations throughout the definition of the bright and dark relaxation times as magnetic field dependent linear combination of 
radiative and non-radiative recombination times. 
We note that, although we chose WS$_2$ as an important example to show interesting results, our model can be straightforwardly extended to WSe$_2$ monolayers by changing the values of parameters, such as the spin-orbit splitting and scattering times. In addition, the theoretical framework can also be applied to Mo-based TMDs after the exchange of the energy level of bright and dark states. For brevity, we leave the cumbersome rate equation in Sec. I of Supplementary Information.  

\section{Conclusion}

We have theoretically investigated the exciton and biexciton dynamics in monolayer WS$_2$ subjected to a tilted magnetic field. We demonstrate that both the intra-valley bright-bright and bright-dark biexcitons posses larger valley splittings than that of bright exciton, namely, -0.46 meV/T and -0.69 meV/T, respectively. In contrast, inter-valley bright-dark biexcitons show an inverted valley splitting of 0.23 meV/T.

The $VP$ of the excitons and biexcitons, on the other hand, is not only determined by the valley splitting, but also the scatterings among the excitonic quasi-particles. We predict that the $VP$ of the intra-valley bright-dark biexciton is strongly enhanced by the magnetic field, and reaches to nearly 50$\%$ at B=65 T. Interestingly, an unusual inversed magnetic response of the $VP$ for the inter-valley bright-dark biexciton is found. A magnetic field can thus rotate the valley pseudospins constructed from these biexcitons in opposite directions, facilitating entanglement of the qubits with controllable phases. Long valley lifetime, large valley splitting, optical controllability and opposite signs of field dependent valley splitting of the bright-dark biexcitons make them appealing candidates for valley qubits for quantum computing.

\section{Acknowledgement}
The authors are grateful to CENAPAD-SP for the provision of computational facilities.
This work was supported by CNPq, FAPDF, FAPESP, and Coordenação de Aperfeiçoamento de Pessoal de Nível Superior - Brasil (CAPES) -
Finance Code 001. H.Z. thanks the financial support from US National Science Foundation (MRI-1229208 and CBET-1510121). H.B. thanks the financial support from FAPESP (grant 2017/23668-8). 

\section{Author contributions}
F.Q. conceived the idea. H.B. and R.V. developed the codes and carried out the simulations; R.V., H.B., and F.L. plotted/designed the figures. H.B., F.Q. and H.Z. wrote the manuscript. All the authors commented on the manuscript.

\section{Competing interests}
The authors declare no competing interests.

\bibliographystyle{unsrt}
\bibliography{refs}

\end{document}